\documentclass{SciPost}

\hypersetup{
    colorlinks,
    linkcolor={red!50!black},
    citecolor={blue!50!black},
    urlcolor={blue!80!black}
}

\usepackage[bitstream-charter]{mathdesign}
\usepackage{dsfont}
\usepackage{mathtools}

\DeclareSymbolFont{usualmathcal}{OMS}{cmsy}{m}{n}
\DeclareSymbolFontAlphabet{\mathcal}{usualmathcal}

\fancypagestyle{SPstyle}{
\fancyhf{}
\lhead{\colorbox{scipostblue}{\bf \color{white} ~SciPost Physics }}
\rhead{{\bf \color{scipostdeepblue} ~Submission }}

\fancyfoot[C]{\textbf{\thepage}}
}

\newcommand{\be}{\begin{equation}}
\newcommand{\ee}{\end{equation}}
\newcommand{\bs}{\begin{subequations}}
\newcommand{\es}{\end{subequations}}

\newcommand{\vmh}{\hat{\vec{m}}}
\newcommand{\vm}{\vec{m}}
\newcommand{\vh}{\vec{h}}
\newcommand{\vn}{\vec{n}}
\newcommand{\vp}{\vec{p}}
\newcommand{\vq}{\vec{q}}
\newcommand{\vz}{\vec{z}}

\newcommand{\nb}{n_\text{B}}
\newcommand{\tg}{\widetilde\gamma}

\begin{document}

\pagestyle{SPstyle}

\begin{center}{\Large \textbf{\color{scipostdeepblue}{
%%%%%%%%%% TODO: Write your article's title here
Landau-Lifshitz damping from Lindbladian dissipation in quantum magnets\\
}}}\end{center}

\begin{center}\textbf{
% Use (full) first name (+ middle name initials) + surname format.
% Separate subsequent authors by a comma, omit comma and use "and" for the last author.
% Mark the corresponding author(s) with a superscript symbol in this order
% \star, \dagger, \ddagger, \circ, \S, \P, \parallel, ...
G\"otz S. Uhrig\textsuperscript{$\star$},
}\end{center}

\begin{center}
Condensed Matter Theory, Department of Physics, TU Dortmund University, 
Otto-Hahn-Stra\ss{}e 4, 44221 Dortmund, Germany
%%%%%%%%%% END TODO: AFFILIATIONS
% Provide email address of corresponding author(s)
\\[\baselineskip]
$\star$ \href{mailto:email1}{\small goetz.uhrig@tu-dortmund.de}
\end{center}

\section*{\color{scipostdeepblue}{Abstract}}
\textbf{\boldmath{%
As of now, the phenomenological classical Landau-Lifshitz (LL) damping of magnetic order is not
 conceptually linked to the quantum theory of dissipation of the Lindbladian formalism which is
unsatisfactory for the booming research on magnetic dynamics.
Here, it is shown that LL dynamics can be systematically derived from Lindbladian dynamics
in a local mean-field theory for weak external fields.
The derivation also extends the LL dynamics beyond the orientation  $\vm/|\vm|$ to the length $|\vm|$
of the magnetization. A key assumption is that the Lindbladian dissipation adapts to the
non-equilibrium $H(t)$ instantaneously to lower its expectation value. 
}}

\vspace{\baselineskip}

%%%%%%%%%% BLOCK: Copyright information
% This block will be filled during the proof stage, and finilized just before publication.
% It exists here only as a placeholder, and should not be modified by authors.
\noindent\textcolor{white!90!black}{%
\fbox{\parbox{0.975\linewidth}{%
\textcolor{white!40!black}{\begin{tabular}{lr}%
  \begin{minipage}{0.6\textwidth}%
    {\small Copyright attribution to authors. \newline
    This work is a submission to SciPost Physics. \newline
    License information to appear upon publication. \newline
    Publication information to appear upon publication.}
  \end{minipage} & \begin{minipage}{0.4\textwidth}
    {\small Received Date \newline Accepted Date \newline Published Date}%
  \end{minipage}
\end{tabular}}
}}
}
%%%%%%%%%% BLOCK: Copyright information

% For convenience during refereeing we turn on line numbers:
%\linenumbers

%\vspace{10pt}
%\noindent\rule{\textwidth}{1pt}
%\tableofcontents
%\noindent\rule{\textwidth}{1pt}
%\vspace{10pt}

% Write your article contents here, starting from first \section.
% An example structure is given below.

\section{Introduction}
\label{sec:intro}

Magnetic dynamics are a topic attracting an enormous interest these days because it is crucial
for making progress in magnonics \cite{barma21}, i.e., for the general idea to use magnetic degrees of freedom for information storage and handling. We stress that this requires to deal with non-equilibrium
physics. Ferromagnetic \cite{chapp07,chuma15} and antiferromagnetic systems \cite{gomon14,baltz18}
as well as spin textures \cite{masel21}
are considered with their respective advantages and disadvantages. 
For substantial progress numerous experimental challenges have to be overcome and well-founded theoretical understanding is indispensable. This requires to capture the relevant Hamiltonian dynamics as well as the relaxation processes. While the former is often known from experimental measurements or from {\it ab initio} calculations, the latter is generically described phenomenologically on a classcial level by the Landau-Lifshitz (LL) equation or the Landau-Lifshitz-Gilbert (LLG) equation 
\cite{landa35,gilbe04,nowak07}.
These equations, though phenomenological, are fundamental to realistic descriptions of magnetism 
away from equilibrium \cite{eriks17}. 

Since spins as the elementary objects of magnetism are quantum entities and in view of the importance of the LL(G) equations it is not surprising that there are intensive efforts to establish a quantum foundation for them
\cite{wiese13,wiese15,sayad15,liu24a,yuan22a,garci24}
uncovering similarities between Lindbladian dynamics for magnons, i.e., 
usual bosons, and results from the LLG equations. Considering magnons as damped
harmonic oscillators, it has been indeed possible to identify a regime where the Lindblad
and the classical LLG equations  coincide in linear order of the deviations from equilibrium 
if a particular bath structure is assumed \cite{yuan22a}. Yet, we would like
to address the issue more generally, i.e., independent of the bath properties as long as
the bath allows for a Lindblad description and independent of a small-angle approximation.
Furthermore, it is pointed out in Ref.\ \cite{garci24} that for an LL(G) description 
to be valid the non-local entanglement between the spins needs to vanish
since the LL(G) behavior is classical. Using a sophisticated Lindblad approach for 
systems of four spins it is concluded that an LL(G)  description may only be 
valid for $S>1$, i.e., for large spin. Yet, very small systems cannot mimic
the behavior of large spin systems in higher dimensions. It is known, that the behavior 
of spins in ordered quantum magnets in higher dimensions can be captured 
well in mean-field approaches because the effect of quantum fluctuations is reduced.
We assume such a high-dimensional system where a local mean-field approach can be used
neglecting non-local entanglement.  The derivation based on such a local mean-field approach 
is mathematically rigorous. We stress that a local mean-field approach still
captures varying lengths of the spin expectation values, i.e., it goes beyond
a purely classical treatment.

Another idea is to establish a quantum mechanical equation 
that results in the established classical equations in the classical limit 
\cite{wiese13,wiese15,liu24a}.
The assumptions necessary to achieve this goal are a strong non-Hermitian Hamiltonian 
\cite{wiese13,wiese15,yu24}
or a self-referential term in the von-Neumann equation where the density matrix appears twice 
\cite{liu24a}. The appearance of a non-Hermitian part has been justified for Bose-Einstein condensates
on heuristic grounds in view of conservation laws \cite{pitae59a,zarem99,shinn20}.
Further studies are called for justifying these assumptions on the basis of physical processes.

In parallel, a number of approaches exists which compute the Gilbert damping parameter obtaining
reasonable values on the basis of time-dependent or non-equilibrium quantum mechanical approaches
addressing the baths explicitly
\cite{antro95,kunes02,capel03,onoda06,ebert11,bhatt12,umets12,sayad15,sakum15,monda23}. 
The resulting master equations, however, do not always amount up to Lindbladian dynamics \cite{basko09}.
Still, a quantum mechanical treatment of the baths must lead to Lindbladian dynamics if its preconditions 
are met: justified Born-Markov approximation 
(bath correlations decay faster than the system's internal dynamics),
and the rotating wave approximation \cite{breue06}. 
Thus, it must be possible to derive the damping in the LL(G) equations from standard quantum mechanics
for open quantum systems in a regime where both descriptions of dissipation are justified.
It is the objective of this paper to provide such a systematically controlled derivation. It is not our goal to
describe damping beyond the validity of a Lindblad approach nor do we aim at incorporating
non-local entanglement to atomistic spin simulations \cite{eriks17,barke19,ander22,berri24}
because we rely on local mean-field theory. But we stress that changes of the spin length as computed
from the squares of the linear spin expectation values are captured by local mean-field
theories.

The Lindblad formalism is a well-established and successful theory for relaxation of quantum systems
under certain conditions. The key
foundation is to consider an open quantum system coupled to a bath \cite{breue06} leading to
the Lindblad master equation including a so-called dissipator. 
If the damping of a magnetic systems 
can be described by the LL(G) equations, which is governed by a single relaxation rate, 
the latter should be derivable from Lindbladian dynamics.
So far, however, Lindblad dynamics and LL(G) dynamics are not linked by a mathematically
rigorous derivation, except that for special systems where their outcomes are the same  \cite{yuan22a}
for small deviations from equilibrium. 
This situation is unsatisfactory:  spins are quantum objects so that the Lindblad approach is applicable.
But the phenomenological Landau-Lifshitz approach is also well established. 
Hence, it suggests itself to ask whether and how these two formalisms are linked.
In this study, we 
try to fill this gap and link Lindbladian dynamics and LL(G) dynamics under certain mild conditions. 
We show that the latter
results from the former in a general limit based on a local mean-field theory, i.e., without
non-local correlations \cite{garci24}. This is justified in higher dimensions. We identify additional terms
in the systematic derivation which describe the change of the length of the magnetization, i.e.,  $|\vm|$.

\section{Brief Recapitulation of Landau-Lifshitz and Lindblad Formalisms}
\label{sec:recap}

As a brief recapitulation, the LL equation reads
\bs
\label{eq:both}
\be
\label{eq:ll}
\frac{d \vmh }{dt} = \vmh\times\vh_0 -\lambda \vmh\times(\vmh\times\vh_0)
\ee
if the Hamilton function reads 
\be
\label{eq:single-spin}
H=-\vh_0\cdot\vmh
\ee
and $\vmh$ is the normalized magnetization
of a spin and $\vh_0$ a fixed external magnetic field with $h_0=g\mu_\text{B} B$. The LLG equation reads 
\be
\label{eq:llg}
\frac{d \vmh }{dt} = \vmh\times\vh_0 -\lambda \vmh\times\frac{d \vmh }{dt}. 
\ee
\es
Both equations are equivalent within ${\cal O}(\lambda^2)$ \cite{laksh84,wiese15} 
because \eqref{eq:llg} can be transformed to 
\be
\label{eq:llg2}
\frac{d \vmh }{dt} = \frac{1}{1+\lambda^2}\left[\vmh\times\vh_0 -\lambda \vmh\times(\vmh\times\vh_0) \right]. 
\ee
Inserting \eqref{eq:llg} in itself and using the 
bac-cab rule $a\times(b\times c)=b(a\cdot c)- c(a\cdot b)$ for $a,b,c\in\mathds{R}^3$
as well as $\vmh\cdot\frac{d \vmh }{dt}=0$ leads to \eqref{eq:llg2}.
Hence the LL and LLG equation stand 
for almost the same dynamics except for a renormalization of the time $t\to (1+\lambda^2)t$.
The difference only appears in second order in $\lambda$.

Lindbladian dynamics for an operator $A$ in the Heisenberg picture is given by the
adjoint quantum Master equation which reads at zero temperature \cite{breue06,yarmo21}
\be
\label{eq:lindblad}
\frac{d\langle A\rangle}{dt} = i \langle[H,A]\rangle + \frac{1}{2}\sum_l 
\gamma_l \Big\langle [B_l,A] B^\dag_l + B_l [A,B^\dag_l]\Big\rangle
\ee
where $H$ is so far a general Hamiltonian, the $\gamma_l$ are the decay rates of the various dissipative channels which are 
defined by the Lindblad operators $B_l$. It is implied that $B_l$ {\it increments}
the energy in $H$ by some energy $\hbar \omega_l>0$ while $B_l^\dag$ decreases the energy 
by the same amount. This energy does not appear in \eqref{eq:lindblad}, bit it is relevant at finite temperatures,
see App.\ \ref{app:finite-temp}.
In the weak-coupling derivation of the Lindblad formalism, one
assumes that the rates $\gamma_l$ are small relative to the time scales of $H$.
But it is shown that the Lindblad structure \eqref{eq:lindblad} does not depend on the
$\gamma_l$ being small \cite{breue06}.
Concretely, we first assume that $H$ is given in Eq.\ \eqref{eq:single-spin}.
At finite temperatures, an additional term occurs which is 
derived and discussed in App.\ \ref{app:finite-temp}. Finite temperature does not change the results
qualitatively, but increases relaxation and prevents that the magnetizations attains its
maximum values.

\section{Single Spin in Time-Dependent Field in Lindblad Formalism}
\label{sec:sspin}

In a first step, we consider a single spin coupled to a static magnetic field which we assume for simplicity to 
point into $z$ direction $\vh_0=h_0 (0,0,1)^\top$ in \eqref{eq:single-spin}. Below, we will extend this assumption. 
The minimal relaxation channel is given by $B=S^-$ with rate $\gamma$, i.e., there is only one decay channel $l=1$
so that we can omit this index to lighten the notation.
Equation  \eqref{eq:lindblad} implies
\bs
\label{eq:fixed-field}
\begin{align}
\frac{d \langle S^z \rangle}{dt} &= 2\gamma S (S- \langle S^z \rangle)
\\
\frac{d \langle S^+ \rangle}{dt} &= -(ih_0 +\gamma S)  \langle S^+ \rangle
\end{align}
\es 
for the spin $S$; with appropriate definitions for longitudinal and transversal relaxation
 these are the Bloch equations \cite{blum12}. Qualitatively, they also 
hold at finite temperatures, see App.\ \ref{app:finite-temp}, {where also the absorption process induced by 
$B^\dag$ contributes leading to terms proportional to the bosonic occupation}. 
The equations \eqref{eq:fixed-field} are exact for $S=1/2$; for general spin
they are valid if the spin state deviates only little from the state in
 the  polarized $z$ direction, i.e., if only the magnetic quantum numbers $S$ and $S-1$ occur.
We will discuss the justification of this approximation below. The solutions of
\eqref{eq:fixed-field} read $\langle S^z \rangle(t)=(\langle S^z \rangle(0)-S)\exp(-2\gamma S t)+S$
and $\langle S^+ \rangle(t)=\exp(-(ih_0+\gamma S)t)\langle S^+ \rangle(0)$. This reflects 
standard longitudinal relaxation combined with damped transverse precession, i.e., the
physically expected result. This behavior is strikingly different from what follows from
Eqs.\ \eqref{eq:both}! Yet this is not the end of the story.

As an important intermediate step we consider the magnetic field $\vh$ 
to be time dependent, i.e., we pass in \eqref{eq:single-spin} from $\vh_0$ to $\vh(t)$
but stick to the single spin case for clarity.
For the Lindbladian approach to be valid, the bath dynamics needs
to be \emph{faster} than the relaxation and the Hamiltonian dynamics (Born-Markov approximation) 
\cite{breue06}. In generic solid state systems, the phonons constitute the bath
and their characteristic energy scale is set by the Debye frequency $\omega_{\rm Debye}$. 
Hence, we assume that $\hbar\omega_{\rm Debye}$ is significantly larger than $\gamma$ and $|\vh|$.
But it is not necessary to specify the physical origin of the bath except that it displays a
rapid internal dynamics. We consider the lack of necessity to 
know more details about the bath an important advantage of the derivation presented
because it renders it general. The rapid internal dynamics allows us to assume that the bath can
adapt to $\vh(t)$ at each instant of time quickly, i.e., the Lindblad operator $B(t)$
adapts to the \emph{instantaneous} magnetic field. Hence the direction of relaxation is
given by $\vh(t)$, i.e., towards aligning the spin with its momentary effective field.
Of course, the instantaneity appears as a strong assumption and future research 
should investigate which effects a relaxation of this assumption has.

Another question arises at this point, namely how the bath ``knows'' about the orientation of
$\vh$. For instance, the above quoted phonons do not couple to the magnetic field and thus
it seems that they cannot induce relaxation towards parallel alignment with the magnetic field.
The answer to this question stems from the fundamental level because the issue of the 
direction of relaxation poses itself already for the case of a constant field $\vh_0$:
fundamental thermodynamics, e.g., Boltzmann's H-theorem,  teaches us that the 
total system consisting of the spin \emph{and} the bath evolves to states of larger
entropy. Concretely, if the bath is at zero temperature as we assume here any excess
energy flows from the system to the bath. This orientation of the relaxation
is incorporated in the above statement that the Lindblad operator $B=S^-$ increases the
energy in $H$ by some positive amount. By extension to slowly varying external fields,
relaxation due to a coupled bath has the preferred direction that energy flows from
the system to the bath at each instant of time.

In App.\ \ref{app:basis} we carry out the systematic rotation of Eqs.\ \eqref{eq:fixed-field}
to a time-dependent reference frame and we obtain for $\vm \coloneqq \langle \vec{S}\rangle$ 
\be
\label{eq:lindblad-fm}
\frac{d\vm}{dt}  = \vm\times\vh + 2\gamma S (S\vn-\vm)-\gamma S
\vn\times(\vn\times\vm),
\ee
where $\vn=\vh/h$ is the time-dependent unit vector in the direction of the 
magnetic field.
In the first term on the right hand side one recognizes the precession and the last term
is a double cross product, but twice with $\vn$ instead of $\vmh$. The
self-reference is missing which is present in Eqs.\ \eqref{eq:both}.

\section{Local Mean-Field Approach and its Numerical Evaluation}
\label{sec:loc-mf}

Here, the next key element enters. We do not attribute the self-reference in Eqs.\ \eqref{eq:both}, i.e.,
the double occurrence of $\vm$ and/or its temporal derivative, to an assumed 
modification of the Schr\"odinger equation, but to a local mean-field approximation 
considering the \emph{ensemble} of spins. This means, we pass
to the translationally invariant Hamiltonian
\be
H=-\sum_{i,j} J_{i,j} \vec S_i \cdot \vec S_j
\ee
with dominating positive couplings so that the fully polarized, ferromagnetic states are the ground states.
The sum  of the couplings
\be
J:= \sum_{j} J_{i,j}
\ee
does not depend on $i$ due to the translational invariance.
In local mean-field theory, it is sufficient to consider an effective single-site
problem, i.e., a single spin in the effective exchange field $\vh_\text{eff} = J\langle \vec S_j\rangle = J \vm$ 
of its interaction partners. Then each spin is subjected
to the external field $\vh_0=h_0(0,0,1)^\top)$ and to this effective exchange field 
 so that the total field reads
\be
\label{eq:eff-field}
\vh = \vh_0 +J\vm
\ee 
and we can resort to the effective single-site Hamiltonian \eqref{eq:single-spin} with $\vh_0\to \vh(t)$
even though we are considering a lattice model.
Note that this approach is easily generalized to arbitrary spin
textures defining a local effective magnetic field $\vh_i\coloneqq \sum_j J_{i,j}\langle \vec S_j\rangle $.
Eventually, one obtains Lindblad equations for each site. If each spin interacts with
a sufficiently large number of other spins or if the spin orientation only changes slowly in space
coarse graining can be performed as well to introduce spin densities. Thus, a number of
extensions is clearly possible.

% figure comparison
\begin{figure}[ht!]
    \centering
        \includegraphics[width=11cm]{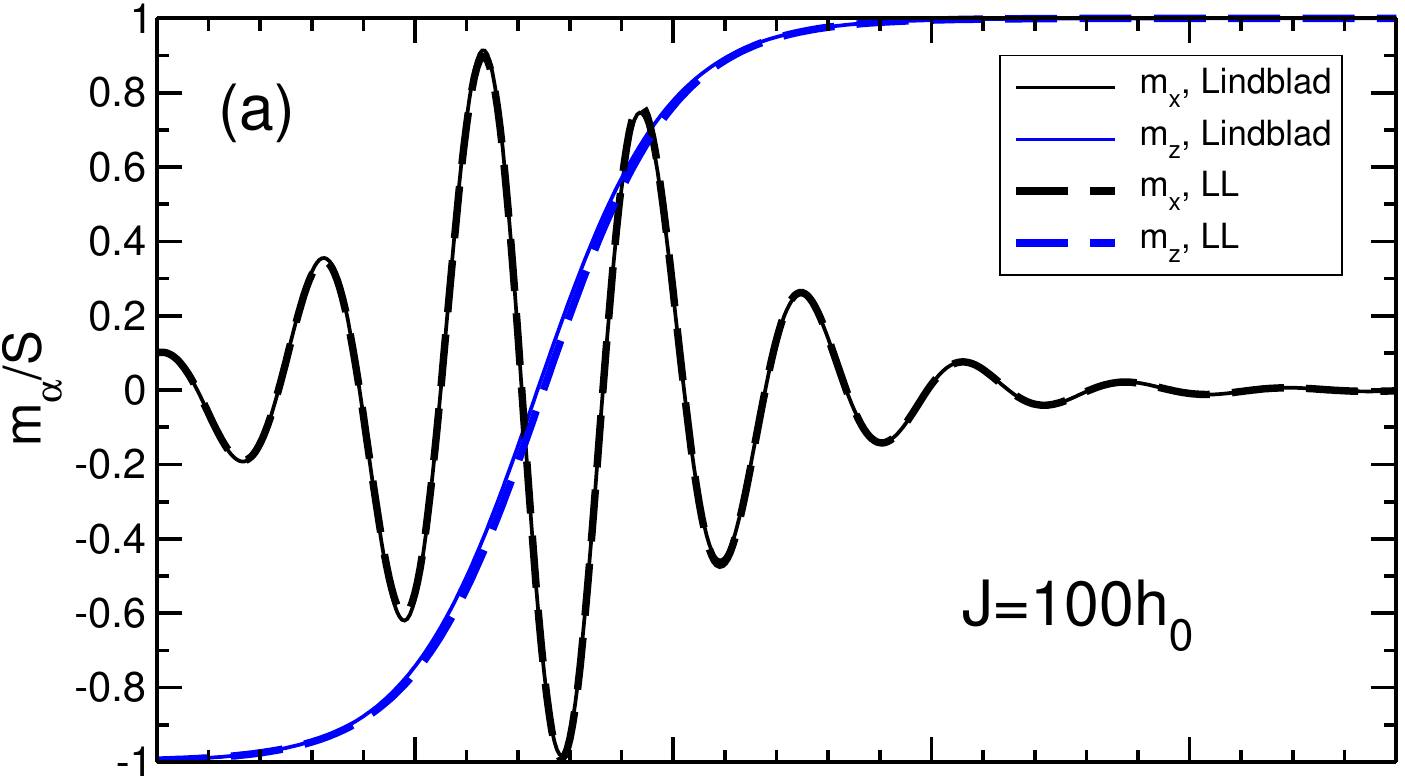}
        \includegraphics[width=11cm]{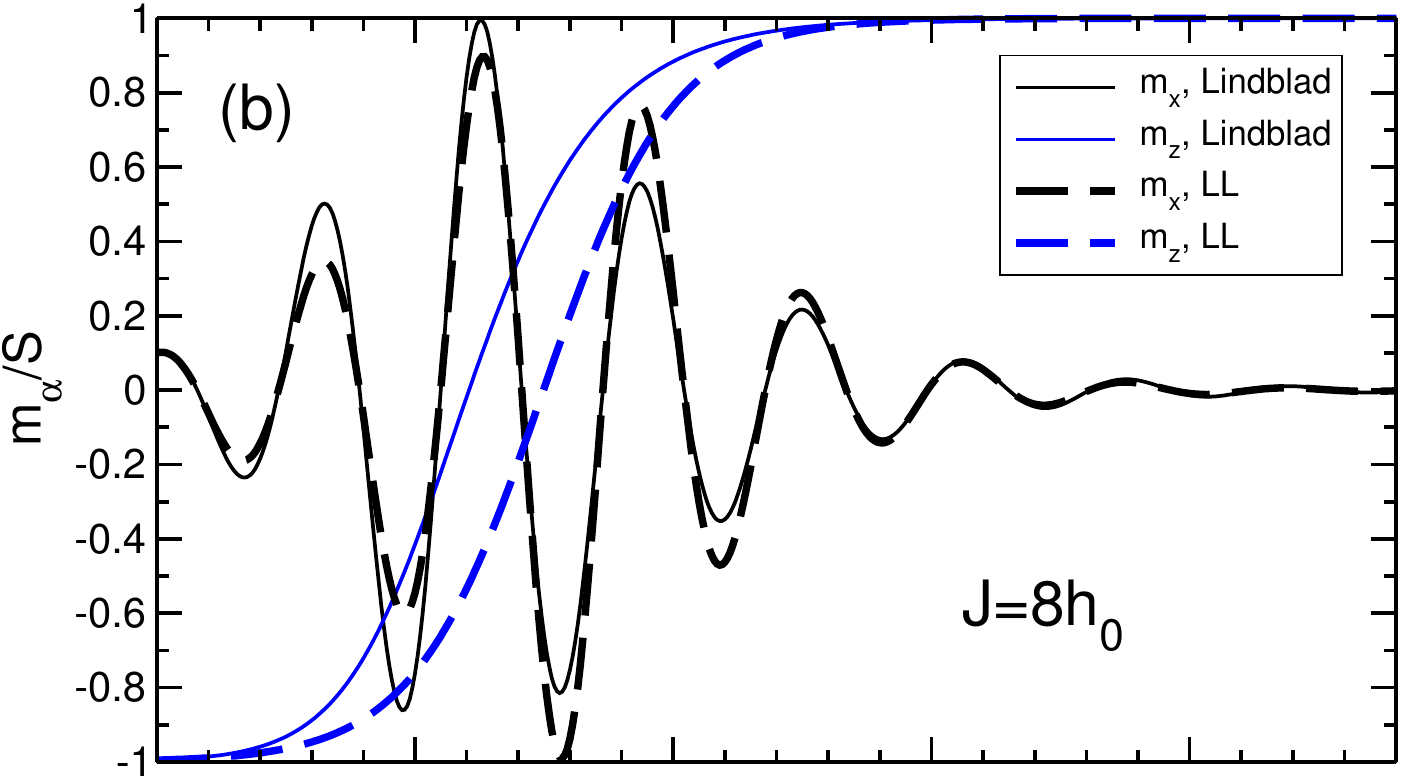}
				\includegraphics[width=11cm]{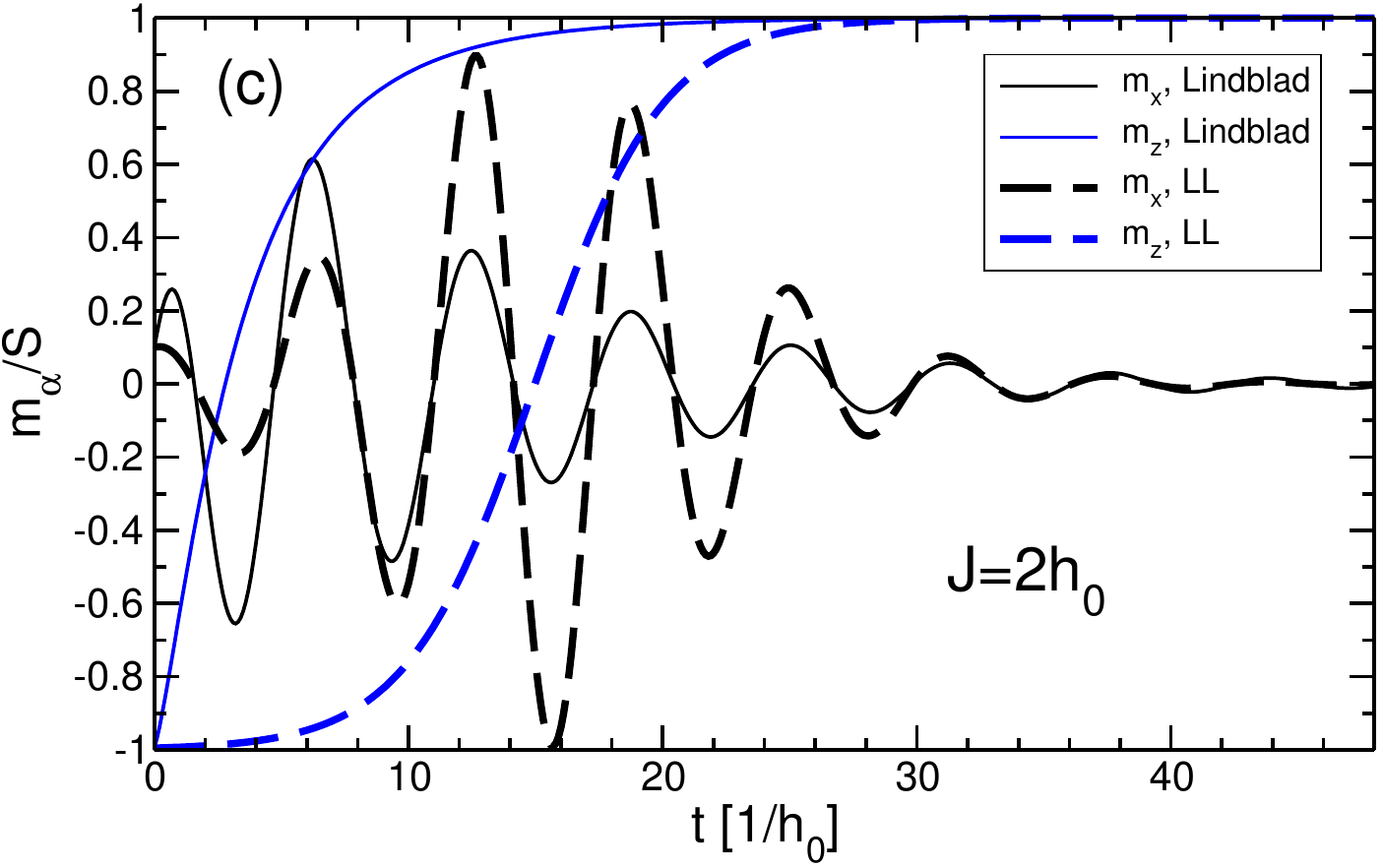}
    \caption{Solutions (solid lines) of Lindbladian
		\eqref{eq:lindblad-fm} with the mean-field self-consistency \eqref{eq:eff-field}
		for the initial length $S$ of the order parameter $\vm$; 
		the initial direction of the magnetization is tilted  by $\pi-0.1$ relative to the $z$ direction.
		The other couplings are $S=1/2$, $\lambda	=0.2$, $J$ is given in the panels, 
		and $\gamma=\lambda J$. The dashed curves represent the solution 
		for the same parameters of the LL equation \eqref{eq:ll}.}
    \label{fig:comparison}
\end{figure}

Inserting \eqref{eq:eff-field} into \eqref{eq:lindblad-fm} yields a closed, 
self-referential set of equations for the magnetization $\vm$. 
Still, it seems to be distinctly different from Eqs.\ \eqref{eq:both},
but this is illusive. A first step to see this is to estimate orders of magnitude.
Generic ferromagnets have a Curie temperature of about $10^3$\,K corresponding roughly
to an internal magnetic field of $10^3$\,T. Externally applied fields range from 1 mT to
10\,T. Thus, we face the situation that $h_0$ is about (and even more than) 
1000 times weaker than the internal field $JS$. This suggests that 
one must focus on the limit of weak external fields, i.e., on $h_0\ll J$.

Figure \ref{fig:comparison} confirms the conjecture that the LL dynamics
is the weak-field limit of the Lindbladian dynamics. The dashed lines show the 
solutions of Eq.\ \eqref{eq:ll} while the solid lines in the same color represent
the solutions of Eq.\ \eqref{eq:lindblad-fm} using \eqref{eq:eff-field}. 
The time axis is displayed in units of $1/h_0$ to show the Larmor precessions clearly.
The relaxation rate $\gamma$ is kept constant relative to the exchange $J$, i.e.,
the parameter $\lambda=\gamma/J$ is fixed.

Clearly, far in the weak field limit for $J/h_0=100$ the agreement is excellent; 
only a minor shift in  the relaxation of the $z$ component is discernible, see panel (a).
Lowering the ratio $J/h_0$ the deviations increase, but the qualitative behavior still
is  close to the LL dynamics, see panel (b). Figure \ref{fig:comparison}(c) shows that for significant
lower ratios the qualitative (!) behavior changes: the $z$ component essentially displays
exponential relaxation as one would expect from \eqref{eq:fixed-field}.
These numerical observations corroborate the above hypothesis that the LL dynamics
results from a Lindblad equations in the limit of weak external fields $h_0$ in comparison
to the strong internal effective fields $J$. If this limit is reached by increasing $J$
while keeping $h_0$ constant, it is important to increase the relaxation rate $\gamma$
as well such that $\lambda=\gamma/J$ remains constant.

At this point, we emphasize that the use of Eq.\ \eqref{eq:eff-field} for large ratios $J/h_0$
implies that the spin orientation is always very close to the direction of the
effective magnetic field $\vh$, not to $\vh_0$. This justifies a posteriori the use of 
the approximation necessary to establish \eqref{eq:fixed-field} for $S>1/2$, i.e.,
it is consistent to use this approximation here.
For $S=1/2$, the differential equations \eqref{eq:fixed-field} are exact anyway.

\section{Analytical Derivation of Generalized Landau-Lifshitz Equation}
\label{sec:analytic}

One could use more numerical data for supporting the above hypothesis.
But it is by far more general to back the numerical agreement by an analytical argument
which captures the dependencies on all parameters.
If the LL dynamics is the weak-field limit of the Lindblad dynamics the LL equation 
must be derivable by a systematic expansion in the external field $h_0$.
Expanding the right hand side of
\eqref{eq:lindblad-fm} in linear order in $h_0$ using $\vn=\vh/h$ and \eqref{eq:eff-field}
yields
\be
\label{eq:analytic1}
\frac{1}{S}\frac{d\vm}{dt} = \frac{\vm}{S}\times \vh_0 + 2\gamma S(\vmh-\vm/S)  
- \frac{\gamma}{J} C\, \vmh\times(\vmh\times\vh_0)
\ee
with $C\coloneqq 2S/|\vm| -1$. the mathematical details of the expansion 
are included in App.\ \ref{app:expansion}. We distinguish here between the magnetization with 
arbitrary length $\vm$ and the normalized one $\vmh$. Inspecting the three terms
on the right hand side of \eqref{eq:analytic1} the first one induces the precession.
It is often quite fast. The second one is also a fast one; 
it could even be faster than the precession depending on the applied field. 
Clearly, it is faster than the third one since we are considering the regime of large $J$. 
The second term ensures that the magnetization
converges to its saturation value $m_\infty=S$ at zero temperature
quickly on the time scale $1/\gamma$. 
At finite temperature, the saturation value $m_\infty$ is lower, see the 
derivation and discussion in App.\ \ref{app:finite-temp}.

The second term in Eq.\ \eqref{eq:analytic1} represents a {\it systematic extension} 
of the LL equations for non-saturated magnetizations still fulfilling all 
required symmetries. Clearly, it goes beyond a purely classical dynamics
because the vector length $|\vm|$ will vary as well; it is a consequence
of the local mean-field approach.
If, however, we start from an initial magnetization that takes its saturation value 
this term has no effect anymore and $|\vm|=m_\infty=S$ holds for all subsequent times. 
This implies $C=1$  and that the time derivative
on the left hand side equals $\frac{d\vmh}{dt}$
as well as the precession term equals $\vmh\times\vh_0$.
 The third term on the right hand side
represents the LL damping term if we set $\lambda=\gamma/J$ as before.
This analytically supports the hypothesis that the LL dynamics 
results from the Lindblad dynamics in the limit of weak external fields.
It is remarkable that in the case of saturation $|\vm|=S$ the time scale
$1/\gamma$ disappears completely and only the time scales $1/h_0$ and $J/(\gamma h_0)$
survive. Such behavior is not uncommon for complex models in particular limits, e.g.,
the rest mass becomes unimportant upon passing from the relativistic to the non-relativistic
energy-momentum relation.
The fact that for a magnetization almost parallel to the external field
the change $\frac{d\vm}{dt}$ becomes very small, even vanishing for parallel orientation,
is an essential feature of the LL equation. Here it follows automatically 
from the Lindblad formalism.

 Thus, we eventually derived the
LL equation \eqref{eq:ll} from Lindbladian dynamics and identified the dimensionless
damping parameter $\lambda$ by the ratio of the relaxation rate $\gamma$ 
to the internal energy scale $J$. If one assumes that the Lindbladian
is derived for weak-coupling limit \cite{breue06}, i.e., the coupling between the system
and the bath is weak it is natural to assume that $\lambda=\gamma/J$ is small.
But, as indicated initially on the Lindblad formalism, this is not mandatory as far as
the Lindblad formalism is concerned. Yet, we restrict our reasoning to $\lambda \le 1$ since
otherwise the LL equation displays unphysical behavior results
indicating that the combination of large relaxation, weak fields, and local mean-field
approximation is not reliable enough. Note that the instability of the LL equation
for $\lambda \le 1$ also applies to the extended equation \eqref{eq:analytic1}
since it is identical to the LL equation for the saturated case $|\vm|=S$.

In addition, we studied larger values of $\lambda$, but still smaller than unity, to 
see whether the LL equation \eqref{eq:ll} or the LLG equation \eqref{eq:llg}
corresponds better to the Lindbladian dynamics \eqref{eq:lindblad-fm}, see App.\ \ref{app:large}.
In accordance with the analytical derivation of the weak-field limit we find
that the LL equation represents the correct limit of weak external fields best.
It its argued that the LLG equation \eqref{eq:llg} is better suited since it 
does not bear the risk of unphysical behavior. It is an intriguing question whether
an extension of the derivation presented here allows one to establish a link
between the Lindblad formalism and the LLG equation. But this issue is beyond the scope
of the present article and thus left to future research.

\section{Relaxation of the Length of the Order Parameter}
\label{sec_length}

Here, we do not refer to the spin length $S$ determining $\langle \vec S^2\rangle =S(S+1)$, but to the length 
of $\vm \coloneqq \langle \vec{S}\rangle$ which we denote by $|\vm|$ or $|m|$ for brevity.
It is interesting that beyond the saturation limit $\vm\ne m_\infty \vmh$ the Eq.\ \ref{eq:analytic1} extends the usual
LL equation \eqref{eq:ll} systematically, see also App.\ \ref{app:finite-temp}.
This can be used to raise  simulations from a classical LL level to the 
level of local mean fields although non-local entanglement is still neglected.
Note that this implies that local quantum aspects are included, namely the possibility
to deal with an order parameter $\vm$ being a spin expectation value which has not a fixed length.

\begin{figure}[ht!]
    \centering
        \includegraphics[width=11cm]{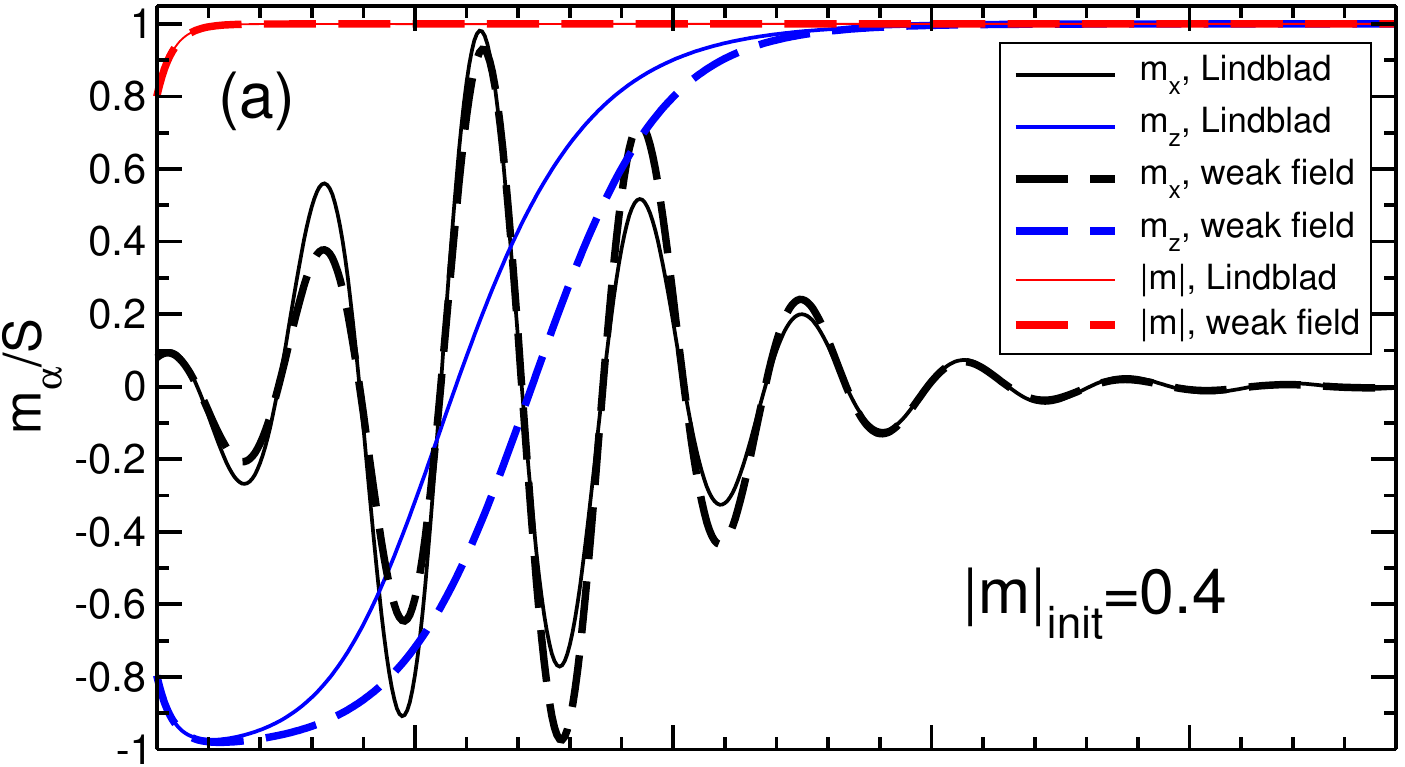}
        \includegraphics[width=11cm]{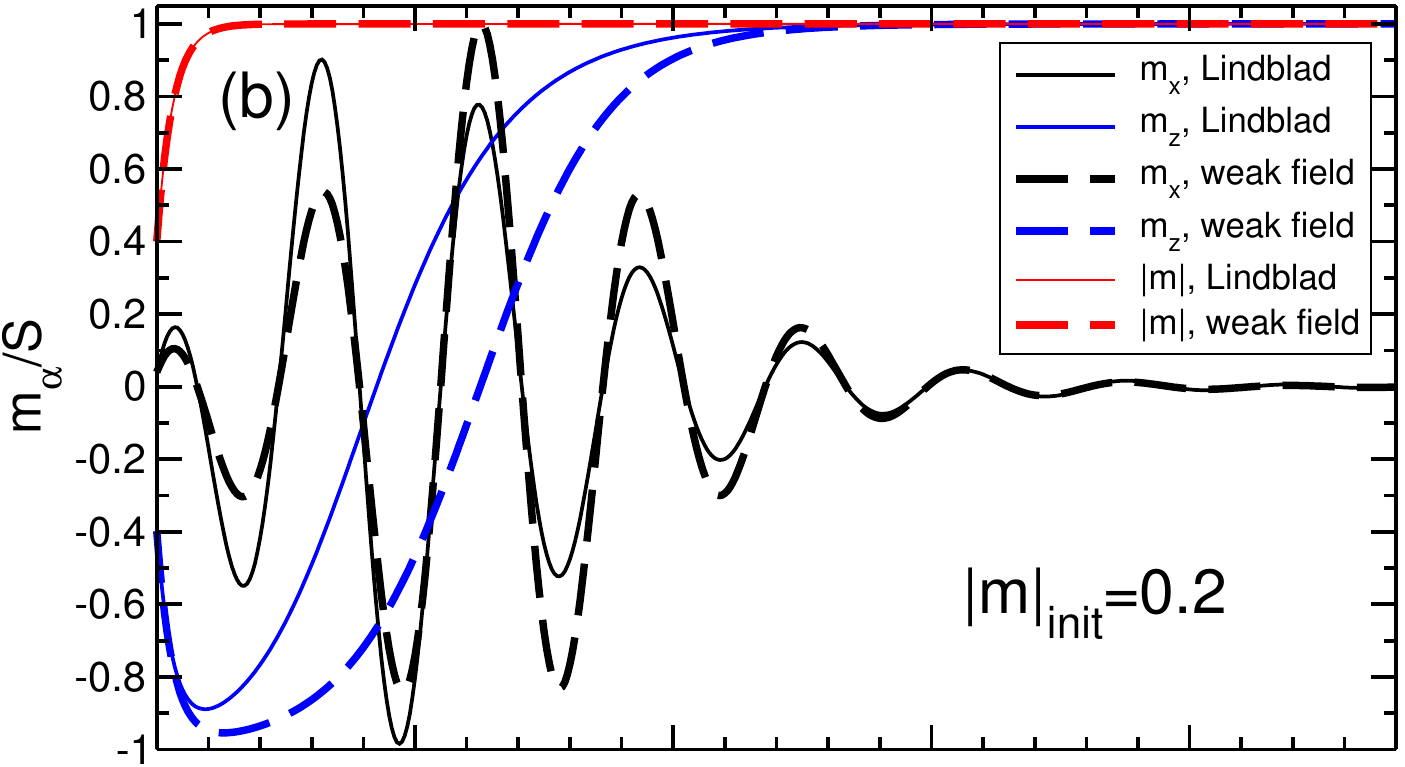}
				\includegraphics[width=11cm]{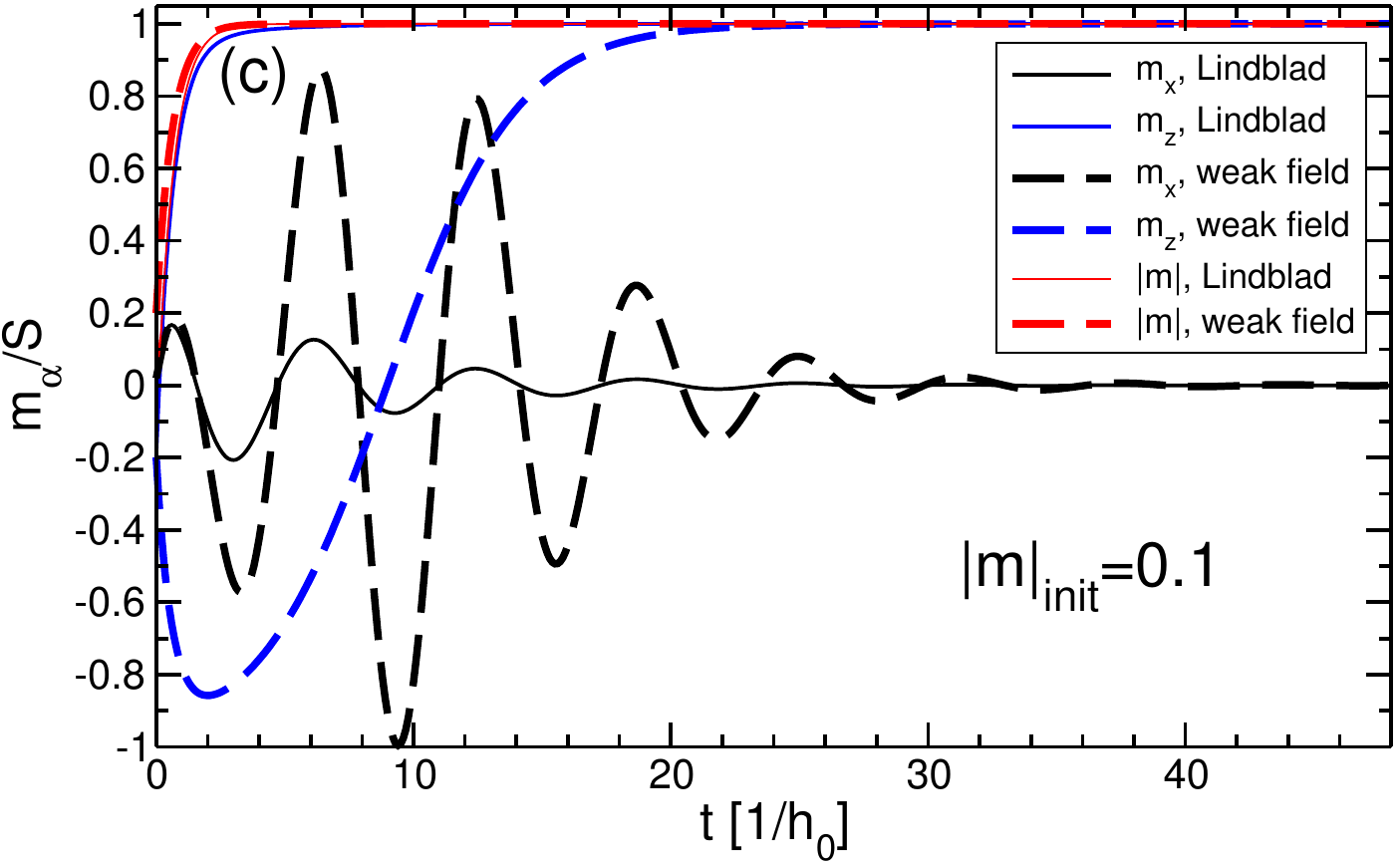}
    \caption{Solutions (solid lines) of Lindbladian
		 \eqref{eq:lindblad-fm} with the mean-field self-consistency \eqref{eq:eff-field}
		for the initial length of $\vm$ given in the panels; 
		the initial direction of the magnetization is tilted  by $\pi-0.1$ relative to the $z$ direction.
		The other couplings are $S=1/2$, $J=8h_0$, $\lambda	=0.2$ so that $\gamma=\lambda J= 8/5.$
		The dashed curves represent the solution 
		for the same parameters of equation \eqref{eq:analytic1} which results 
		in the weak-field limit $h_0/J\to 0$ in linear order.  }
    \label{fig:length}
\end{figure}

Figure \ref{fig:length} extends the calculations shown in Fig.\ 1b which started from the saturated value $|\vm|=S$
to smaller values corresponding to an initially smaller value of the order parameter $|\vm|<S$. It becomes evident
that the relaxation of the length occurs rapidly in accordance with the larger values of $\gamma=\lambda J$. 
Interestingly, it appears that first the length of the order parameter enters into saturation before its orientation
changes significantly. This is true even if its orientation is antiparallel to the external field, note the initial
behavior of $m_z$ in Figs.\ 2a and 2b where the initial order parameter is still non-negligible.
In this regime, the approximate leading-order equation \eqref{eq:analytic1} still captures the dynamics qualitatively,
similar to what we saw in Fig.\ 1b for $|\vm|=S$. The agreement becomes significantly better if we studies smaller values of
$h_0/J$.

But Fig.\ 2c illustrates that the approximate equation fail to reflect the true Lindbladian
dynamics, even qualitatively, 
if the inital order parameter is too small. This was to be expected because the length $|\vm|$ occurs in
\eqref{eq:analytic1} in the denominator of the relaxational term formed by the double cross product.
Here, one may wonder whether the coefficient $C$ in \eqref{eq:analytic1} can change sign. This would have
important physical consequences because the double cross product with negative sign ensures that the orientation
aligns with the external field in the long run $t\to\infty$. A different sign would spoil this wanted
relaxational behavior. But from the inequality $|\vm|\le S$ which is ensured for the quantum
mechanical expectation value $\langle \vec S\rangle$, it is evident that $C\ge 1$. Still, it cannot
be excluded rigorously that the dynamics in \eqref{eq:lindblad-fm} or \eqref{eq:analytic1} lead to an
overshooting $|\vm|\ge S$. But none of our test runs starting from physical reasonable initial conditions
showed any sign of such overshooting.

In summary, Eq.\ \eqref{eq:lindblad-fm} using \eqref{eq:eff-field} constitutes a justified extension
of the LL equation \eqref{eq:ll} including the dynamics of the length of the vector valued order parameter.

\section{Conclusions and Outlook}
\label{sec_conclusio}

In summary, up to now there was a conceptual gap between the established
Lindbladian theory for dissipation in open quantum systems and the phenomenological
theory of Landau, Lifshitz, and Gilbert for the damping of the temporal behavior
of magnetic order. In view of the humongous interest in magnetic dynamics this
represented a  caveat. 
So far, the differences of the two approaches were emphasized, for instance the
change of entropy in Lindblad treatments, see e.g.\ Ref.\ \cite{uhrig20}, 
and the conservation of entropy in the assumed quantum analogues
of the LL(G) equations  \cite{wiese13,liu24a}. Other papers succeeded in linking
an LL and a Lindblad description for specific models. In the present study, we filled the
conceptual gap and reconciled the  quantum description in terms of a Lindbladian
master equation and the phenomenological classical LL(G) description on a fundamental level. 
In addition, a systematic
extension for the evolution of the length of the expectation value of $\vec{S}$
has been found. This underlines that the Lindblad approach, though based on
a local mean-field theory, reaches beyond a merely classical description.

Three key points constitute the important ingredients: 
(i) the Lindbladian relaxation is constantly adapted to the instantaneous
Hamilton operator and its energies, i.e., the relaxation always acts such as to minimize
the instantaneous energy; (ii) the self-reference in the LL(G) equations is
explained not by ad hoc modifications of the Schr\"odinger equation, but by 
a local mean-field approximation; (iii) finally the internal energy scales of the quantum
system need to be much larger than the external ones, here $|J|\gg |h_0|$.
Key point (i) requires that the internal bath dynamics is faster than the system's dynamics.
This is a standard assumption in the derivation of the Lindblad master equation.
These ideas allowed us to derive the Landau-Lifshitz equation analytically from
Lindbladian dynamics and to confirm this by numerical calculations.
We did not observe a trace of Landau-Lifshitz-Gilbert dynamics for larger values of
$\lambda=\gamma/J\le 1$.

The possibility to derive LL dynamics from Lindbladian dynamics provides the
general context for computing damping parameters from elaborate theories 
describing the reservoirs explicitly which are responsible for the relaxation of quantum
system \cite{antro95,kunes02,capel03,onoda06,ebert11,bhatt12,umets12,sayad15,sakum15}. 
The above ideas can also be applied directly to quantum magnets beyond
ferromagnets as along as they are treated by a local mean-field theory. An important
example are quantum antiferromagnets gaining more and more interest for the
prospect of faster data handling \cite{gomon14,behov23} or general spin
textures \cite{masel21}. We stress that quantum fluctuations of the spin ensemble itself
are not captured; they would require advanced mean-field techniques \cite{bolsm23,khudo24a}
or modified approaches to atomistic spin dynamics \cite{eriks17,barke19,ander22,berri24}. 

Another extension can consist in using Eq.\ \eqref{eq:analytic1} instead of the LL equation
in atomistic spin simulations \cite{nowak07,coffe12,nishi15,eriks17}. Such approaches
are also employed to describe systems of Josephson junctions \cite{guarc20,guarc21,mazan24}.
This would allow for taking changes in the spin length into account, promoting the
classical calculation to a local mean-field one if the external fields are weak
with respect to the internal ones. If the weak-field limit is not justified one
can equally well resort to Eq.\ \eqref{eq:lindblad-fm} which is based on a local
mean-field approach, but does not require a particular energy hierarchy between
internal and external magnetic energies, except that the bath is assumed to be internally fast. 
For including thermal fluctuations by
stochastical noise terms, the conditions for reaching equilibrium need to be 
verified \cite{nishi15}.

Recently, high frequencies in the spin dynamics of ferromagnets 
have been predicted and experimentally 
identified which are attributed
to spin inertia and a resulting nutation term in the Landau-Lifshitz-Gilbert
equations \cite{ciorn11,thoni17,cherk20,neera21}. Mathematically, a second time derivative
of the magnetization is included. This poses the question whether such a term
can result from a Lindblad formalism by extending the above reasoning.

The success of the presented calculations clearly suggests that Lindbladian
dissipation captures the physics of relaxation very well. The above key idea (i) that
it must be constantly adapted to the instantaneous Hamiltonian
in time-dependent, non-equilibrium situations is generally applicable if the energy scales
are such that the internal bath dynamics are the fastest. Thus, the reported findings 
may pave the way to studies of relaxation in a plethora of non-equilibrium quantum
systems.

\section*{Acknowledgements}
The author is thankful to Timo Gr\"a\ss{}er, Asliddin Khudoyberdiev, and Joachim Stolze 
for useful discussions.

% TODO: include funding information
\paragraph{Funding information}
This research has been supported by the Deutsche Forschungsgemeinschaft
(DFG) in projects UH 90-14/1 and UH 90-14/2 as well as by the Stiftung Mercator
in project KO-2021-0027.

\begin{appendix}
\numberwithin{equation}{section}

\section{Derivation of Lindbladian Dynamics for Time-Dependent Magnetic Field} %former SM 
\label{app:basis}

The aim is to derive Eq.\ \eqref{eq:lindblad-fm}. This implies that we carry out
a rotation in spin space to account for the magnetic field varying in time.
Concomitantly, the Lindblad operator varies in time as well since it is assumed
to relax the spin at each instant of time according to the instantaneous direction 
of the field, i.e.,
such that the spin approaches parallel orientation to the magnetic field.

We choose $\vn\coloneqq \vh/h$ and an
orthogonal unit vectors $\vp$ so that they for a orthonormal basis of $\mathds{R}^3$ 
together with $\vq\coloneqq\vn\times \vp$. Their completeness is expressed by
\be
\label{eq:identity}
\vn\vn^\top + \vp\vp^\top  + \vq\vq^\top  = \mathds{1}_3
\ee
where $\mathds{1}_3$ is the $3\times3$ identity matrix. Differentiation of the
completeness with respect to time yields
\be
\label{eq:vanish}
\frac{d\vn}{dt}\vn^\top + \vn\frac{d\vn}{dt}^\top
+  \frac{d\vp}{dt}\vp^\top + \vp\frac{d\vp}{dt}^\top
+ \frac{d\vq}{dt}\vq^\top + \vq\frac{d\vq}{dt}^\top= 0
\ee
which we will use below. For a concise notation we introduce the complex
vector
\be
\vz \coloneqq \vp+ i \vq \in\mathds{C}^3.
\ee
Then we can compute the longitudinal and transverse spin expectation value
using $\vm= \langle \vec{S} \rangle$
\bs
\label{eq:components}
\begin{align}
S_\parallel &\coloneqq  \vn\cdot \vm
\\
S_+ &\coloneqq  \vz\cdot \vm.
\end{align}
\es
These values acquire their time dependence from 
Eqs.\ \eqref{eq:fixed-field} 
and from the time dependence of $\vn$ and $\vz$
\bs
\label{eq:diff-components}
\begin{align}
\frac{d S_\parallel}{dt} &= \gamma 2S(S-S_\parallel) + \frac{d\vn}{dt}\cdot \vm
\\
\frac{d S_+}{dt} &=   -(ih+\gamma S)S_+ + \frac{d\vz}{dt}\cdot \vm.
\end{align}
\es

Reconstructing $\vm$ from the two components defined in \eqref{eq:components} yields
\bs
\begin{align}
\vm &= \vn S_\parallel +\frac{1}{2}\vp\, (S_++S_+^*) -\frac{i}{2}\vq\,(S_+ -S_+^*)
\\
&=  \vn S_\parallel + \Re(\vz^{\,*} S_+).
\label{eq:vm1}
\end{align}
\es
Computing the derivative of $\vm$ from \eqref{eq:vm1} and \eqref{eq:diff-components}
yields
\bs
\begin{align}
\frac{d \vm}{dt} &= 
\frac{d \vn}{dt} (\vn\cdot\vm) + \vn\Big(\frac{d \vn}{dt} \cdot \vm\Big) 
\label{eq:line1}\\
&\quad + 
\Re\Big(\frac{d\vz^{\,*}}{dt} (\vz\cdot\vm)+ \vz^{\,*}(\frac{d\vz}{dt}\cdot\vm) \Big)
\label{eq:line2} \\
& + \gamma 2S\vn (S-\vn\cdot\vm) -\Re \big((ih+\gamma S) \vz^{\,*}(\vz\cdot\vm)\big).
\end{align}
\es
The lines \eqref{eq:line1} and \eqref{eq:line2} cancel due to \eqref{eq:vanish}.
Using 
\be
\Re(\vz^{\,*}\vz^\top) = \vp\vp^\top + \vq\vq^\top = \mathds{1}_3 -\vn\vn^\top,
\ee
where the second equation results from \eqref{eq:identity}, as well as 
\bs
\begin{align}
\Im (\vz^{\,*}(\vz\cdot\vm)) &= \vp(\vq\cdot\vm)-\vq(\vp\cdot\vm)\\
 & = \vm\times(\vp\times\vq) = \vm \times\vn
\end{align}
\es
leads us to 
\be
\frac{d \vm}{dt} = \vm\times \vh +\gamma 2S^2 \vn -\gamma S\vn(\vn\cdot\vm) - \gamma S \vm.
\ee 
To the right hand side we add and subtract $\gamma S\vm$ so that we can use 
$\vm - \vn(\vn\cdot\vm) = - \vn\times(\vn\times\vm)$ to obtain
\be
\label{eq:sought}
\frac{d \vm}{dt} = \vm\times \vh +\gamma 2S(S\vn-\vm) -\gamma S\vn\times(\vn\times\vm).
\ee 
This is the sought equation \eqref{eq:lindblad-fm}.

\section{Linear Order in the External Field $h_0$}\ %former SM 
\label{app:expansion}

Here we carry out an expansion of Eq.\ \eqref{eq:lindblad-fm}, identical to \eqref{eq:sought}, 
in orders of the external magnetic field $h_0$. There is no zeroth order so that we aim at the linear order.
The first term in \eqref{eq:sought} takes the form
\be
\label{eq:term1}
\vm\times \vh = \vm \times \vh_0
\ee
because $\vm\times\vm=0$; this is linear in $h_0$. In the second term, we
have to use 
\be
h^2= J^2|\vm|^2 + 2J \vh_0\cdot \vm  + h_0^2
\ee
to reach
\bs
\begin{align}
\vn &= \frac{\vh}{h}
\\ &= \frac{\vh_0}{J|\vm|} + \vmh\Big(1-\frac{\vh_0 \cdot \vmh}{J|\vm|}\Big) +
{\cal O}(h_0^2)
\\
&= \vmh +\frac{1}{J|\vm|}(\vh_0 -\vmh(\vmh\cdot\vh_0)) + {\cal O}(h_0^2)
\\
&= \vmh - \frac{1}{J|\vm|} \vmh\times(\vmh\times\vh_0) + {\cal O}(h_0^2)
\label{eq:term2}
\end{align}
\es
where we use the normalized magnetization $\vmh=\vm/|\vm|$. The third term in 
\eqref{eq:sought} simplifies according to 
\bs
\begin{align}
\vn\times(\vn\times\vm) &= \frac{\vh\times(\vh\times\vm)}{h^2}
\\
&= \frac{\vh\times(\vh_0\times\vm)}{h^2}
\\
&= \frac{J\vm\times(\vh_0\times\vm)}{J^2|\vm|^2}+ {\cal O}(h_0^2)
\\
&= - \frac{\vmh\times(\vmh\times\vh_0)}{J}+ {\cal O}(h_0^2).
\label{eq:term3}
\end{align}
\es
Adding the results \eqref{eq:term1}, \eqref{eq:term2}, and \eqref{eq:term3}
with the appropriate prefactors yields
\be
\frac{d \vm}{dt} = \vm\times \vh_0+ \gamma 2S(S\vmh -\vm) -
\frac{\gamma S}{J} \Big(2\frac{S}{|\vm|} -1\Big)\vmh\times(\vmh\times\vh_0).
\ee 
Dividing by $S$ yields the claimed equation \eqref{eq:analytic1}.

\section{Effect of Large Landau-Lifshitz Damping}
\label{app:large}

\begin{figure}[ht!]
    \centering
        \includegraphics[width=11cm]{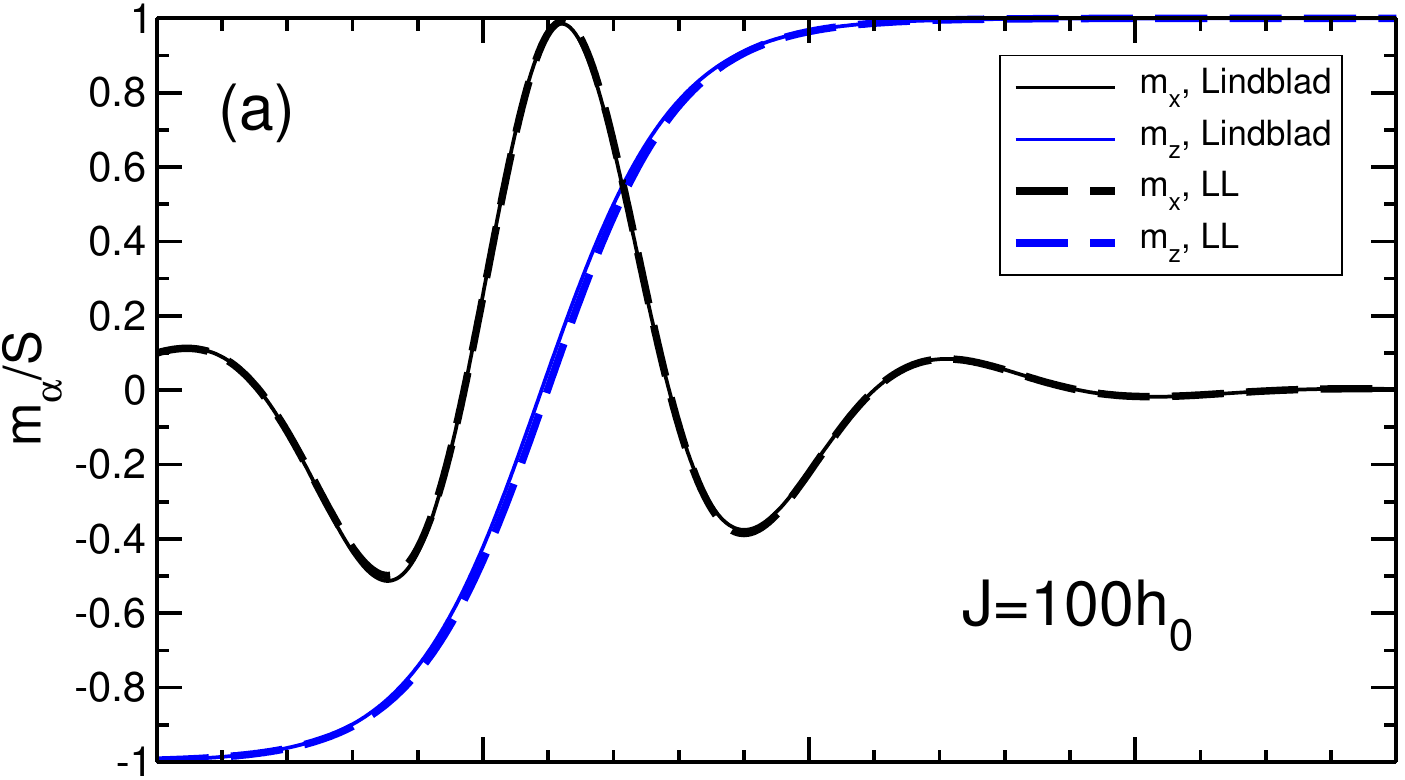}
        \includegraphics[width=11cm]{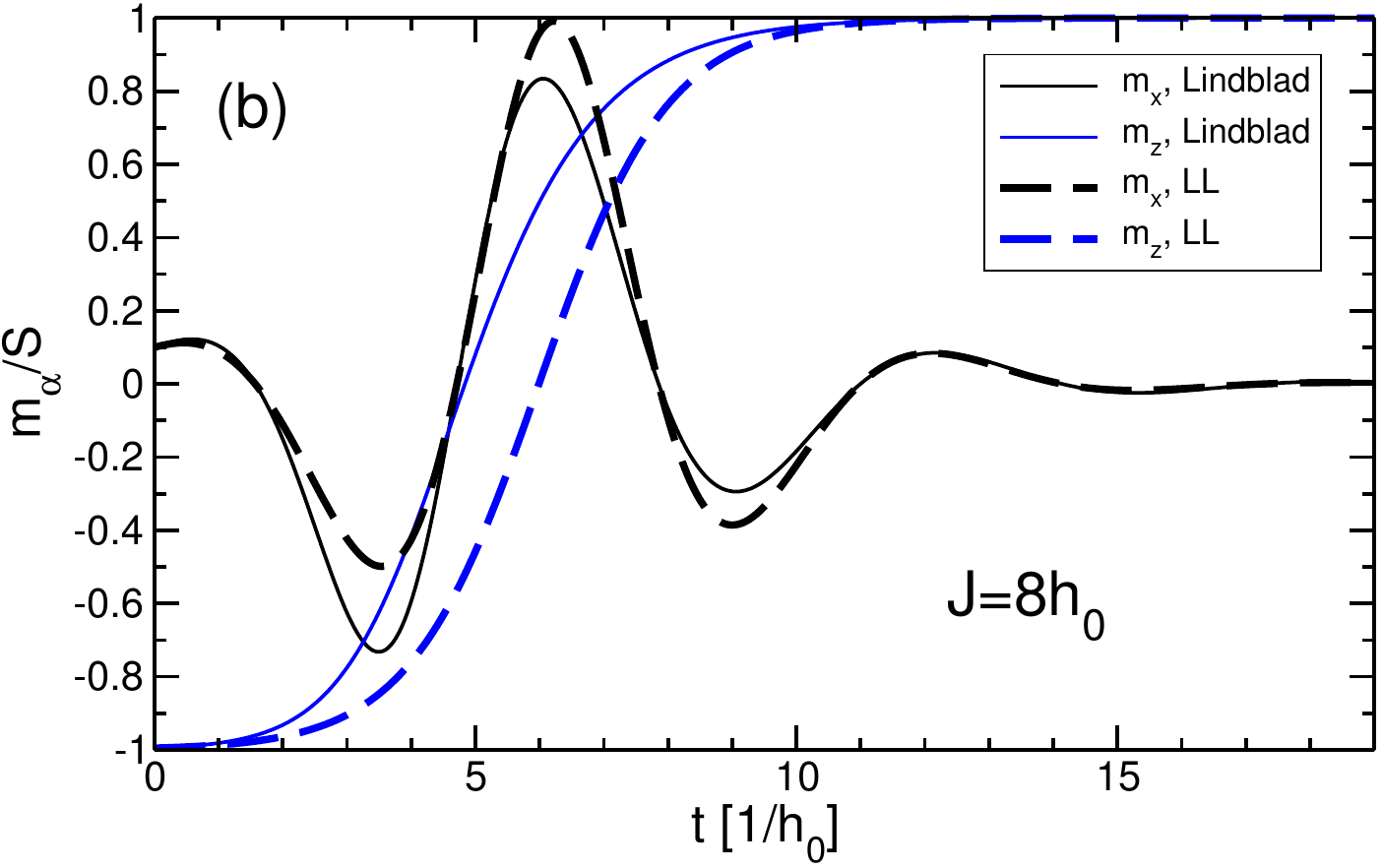}
    \caption{Solutions (dashed lines) of the Landau-Lifshitz equation \eqref{eq:ll} for
		$\lambda=\gamma/J=0.2$ and $S=1/2$. 
		Solutions (solid lines) of the Lindbladian
		equation \eqref{eq:lindblad-fm}.  The initial direction of the magnetization of length $S$
		is tilted  by $\pi-0.1$ with respect to the $z$ direction.}
    \label{fig:large-lambda}
\end{figure}

The figures in the article and the analytical calculation 
presented above in the previous section indicate that the LL equation \eqref{eq:ll}
captures the limit of weak external fields of the Lindbladian dynamics best. Yet, the damping value
used in the numerical computation, $\lambda=0.2$, is fairly small and any possible effect 
resulting from the LLG equation \eqref{eq:llg} 
on the dynamics would be only $4\%$, see Eq.\ \eqref{eq:llg2}.
Hence, we study a larger values of $\lambda$. Yet we stick to values $\lambda<1$ since
it is known that the LL equation \eqref{eq:ll} yields unphysical behavior for $\lambda>1$.
Moreover, we only need to identify possible quadratic effects which distinguish
Eq.\ \eqref{eq:ll} from Eq.\ \eqref{eq:llg2}.
Thus, to corroborate the observation that the Lindbladian dynamics reduces
to LL dynamics rather than to LLG dynamics we study the case $\lambda=0.5$ in 
Fig.\ \ref{fig:large-lambda}. 

% figure large lambda

Of course, the damping takes place faster by a factor of $2.5$ than in 
Fig.\ \ref{fig:comparison}.
In summary, however, 
we find again that the agreement for large $J$ in  panel (a) is almost perfect while
deviations become discernible for intermediate $J$ in  panel (b). But the qualitative behavior
remains the same. We do not observe any shift in the oscillations of the $S^x$ component
relative to the oscillations in the corresponding LL curve. Hence, we
do not see any sign of a renormalization of the precession term. From the LLG equation 
\eqref{eq:llg2},
one would have expected a $25\%$ effect. Also the dynamics of the $S^z$ remains very much the
same: it compares similarly to the LL dynamics as we observed in Fig.\ \ref{fig:comparison}.
Definitely, there is no $25\%$ effect. Hence, the weak-field limit is given by the LL
dynamics \eqref{eq:ll} rather than by the LLG dynamics \eqref{eq:llg}.

\section{Effect of Finite Temperature on the Dissipation} %former SM 
\label{app:finite-temp}

At finite temperature, the dissipator denoted in Eq.~\eqref{eq:lindblad} must
be {modified \cite{yarmo21} by adding the term}
\be
T_1 \coloneqq \frac{1}{2}\sum_l \gamma_l \nb(\hbar \omega_l) 
\Big\langle \big[[B_l,A] B^\dag_l\big] + \big[B_l [A,B^\dag_l]\big]\Big\rangle
\ee
where $\nb$ is the bosonic occupation number at inverse temperature  $\beta$
given by $\nb=(\exp(\beta \hbar\omega_l)-1)^{-1}$ as usual. 
{We stress that $T_1$ includes the effects of the Lindblad operators $B_l$
inducing decay as well as the effects of the Lindblad operators $B_l^\dag$
inducing absorption processes.}
Here, $\hbar\omega_l$ is the energy increment induced by the Lindblad operator $B_l$. 
Specifying to the considered case of a spin in a magnetic field, 
we only have one $l=1$ and $B_1=S^-$ with $\hbar\omega_1=J|\vm|$. Carrying out the calculation
we find the extended version of Eq.\ (4)
\bs
\label{eq:fixed-field-temp}
\begin{align}
\frac{d \langle S^z \rangle}{dt} &= 2\gamma S (S- (1+\nb/S)\langle S^z \rangle)
\\
\frac{d \langle S^+ \rangle}{dt} &= -(ih_0 +\gamma S(1+\nb/S))  \langle S^+ \rangle.
\end{align}
\es 
To account for the changes, we define $\tg\coloneqq \gamma (1+\nb/S)$ allowing us 
to rewrite the above equations as 
\bs
\label{eq:fixed-field-temp2}
\begin{align}
\frac{d \langle S^z \rangle}{dt} &= 2\tg S (S-\langle S^z \rangle)- 2\tg S \delta
\\
\frac{d \langle S^+ \rangle}{dt} &= -(ih_0 +\tg S)  \langle S^+ \rangle
\end{align}
\es 
with $\delta \coloneqq \nb/(1+\nb/S)$. These changes modify Eq.\ (5) to
\be
\label{eq:lindblad-fm-tmp}
\frac{d\vm}{dt}  = \vm\times\vh + 2\tg S ((S-\delta)\vn-\vm)-\tg S \vn\times(\vn\times\vm)
\ee
and in linear order in $h_0$ Eq.\ (7) becomes
\begin{align}
\label{eq:analytic1-temp}
\nonumber
\frac{1}{S}\frac{d\vm}{dt} &= \frac{\vm}{S}\times \vh_0 
+ 2\tg S\left(\frac{\vmh}{1+\nb/S}-\frac{\vm}{S}\right)
\\ & \quad  
 - \frac{\tg}{J} C_T \, \vmh\times(\vmh\times\vh_0)
\end{align}
with $C_T\coloneqq 2S/[|\vm|(1+\nb/S)] - 1$.
Note that the length $m_\infty$ of the magnetization in the stationary regime 
is reduced from $S$ at zero temperature to $S/(1+\nb/S) < S$ at finite temperature.
For $S=1/2$ this is the same result as in a local mean-field theory. For larger values
of $S$ only the leading influence of $T>0$ is captured because Eq.\ (4) is not
exact for $S>1/2$, but only a good approximation close to full polarization. 
Of course, this can be remedied if required, for instance by modifications
in an atomistic spin simulation \cite{eriks17,barke19,berri24}.

Inspecting the equations \eqref{eq:lindblad-fm-tmp} and \eqref{eq:analytic1-temp}
the main difference relative to the dynamics at zero temperature is an enhanced relaxation
rate $\tg>\gamma$ and a difference in the long-time limit $m_\infty$ of the absolute value of the
 magnetization $\vm$. As expected, finite temperature enhances the relaxation and reduces
$m_\infty$ from  its value at zero temperature. Hence, if one starts from the
long-time limit, which is also the equilibrium value in mean-field theory for  $S=1/2$,
the equations are the same as at zero temperature except for the change in the relaxation rate.
In particular, we find that the Landau-Lifshitz equation can be deduced from the 
Lindbladian under mild and plausible assumptions. This extends the
key result  of the main text to finite temperatures.

%\section{About references}
%\begin{verbatim}
%\bibliography{ll-lindblad}
%\end{verbatim}

\end{appendix}


% Provide your bibliography here. You have two options:

%%% FIRST OPTION
% Write your entries here directly, following the example below, including:
% Author(s), Title, Journal Ref. with year in parentheses at the end, followed by the DOI number.

%\begin{thebibliography}{99}
%\bibitem{1931_Bethe_ZP_71} H. A. Bethe, {\it Zur Theorie der Metalle. i. Eigenwerte und Eigenfunktionen der linearen Atomkette}, Zeit. f{\"u}r Phys. {\bf 71}, 205 (1931), \doi{10.1007\%2FBF01341708}.
%\bibitem{arXiv:1108.2700} P. Ginsparg, {\it It was twenty years ago today... }, \url{http://arxiv.org/abs/1108.2700}.
%\end{thebibliography}

% Use your bibtex library, formatted by the SciPost style file.
%\bibliography{ll-lindblad.bib}

%%%%%%%%%% END TODO: BIBLIOGRAPHY

\end{document}